\documentclass[prl,twocolumn,letterpaper,10pt,twoside,tightenlines,nofootinbib,showpacs]{revtex4}
\usepackage{amsmath,amsfonts,amssymb,latexsym}
\usepackage{graphicx,color}
\usepackage[sort&compress]{natbib}

\begin{document}

\newcommand{\eg}{{\it e.g.}}
\newcommand{\cf}{{\it cf.}}
\newcommand{\etal}{{\it et. al.}}
\newcommand{\ie}{{\it i.e.}}
\newcommand{\be}{\begin{equation}}
\newcommand{\ee}{\end{equation}}
\newcommand{\bea}{\begin{eqnarray}}
\newcommand{\eea}{\end{eqnarray}}
\newcommand{\bef}{\begin{figure}}
\newcommand{\eef}{\end{figure}}
\newcommand{\bce}{\begin{center}}
\newcommand{\ece}{\end{center}}
\newcommand{\red}[1]{\textcolor{red}{#1}}

\newcommand{\dd}{\text{d}}
\newcommand{\ii}{\text{i}}
\newcommand{\lsim}{\lesssim}
\newcommand{\gsim}{\gtrsim}
\newcommand{\RAA}{R_{\rm AA}}

\title{Collectivity of $J/\psi$ Mesons in Heavy-Ion Collisions}

\author{Min~He$^{1}$, Biaogang~Wu$^2$, and Ralf~Rapp$^2$}
\affiliation{$^1$Department of Applied Physics, Nanjing University of Science and Technology, Nanjing~210094, China}
\affiliation{$^2$Cyclotron Institute and Department of Physics and
Astronomy, Texas A\&M University, College Station, Texas 77843-3366, U.S.A.}

\date{\today}

\begin{abstract}
The production of $J/\psi$ mesons in heavy-ion collisions at the Large Hadron Collider is believed to be dominated by the recombination of charm
and anti-charm quarks in a hot QCD medium. However, measurements of the elliptic flow ($v_2$) of $J/\psi$ mesons in these reactions are not
well described by existing calculations of $J/\psi$ recombination for transverse momenta $p_T\gsim 4$\,GeV. We revisit these calculations
in two main aspects. Employing the resonance recombination model, we implement distribution functions of charm quarks transported
through the quark-gluon plasma using state-of-the-art Langevin simulations and account for the space-momentum correlations of the diffusing
charm and anti-charm quarks in a hydrodynamically expanding fireball. This extends the relevance of the recombination processes
to substantially larger momenta than before. We also revisit the suppression of primordially produced $J/\psi$'s by propagating them
through the same hydrodynamic medium, leading to a marked increase of their $v_2$ over previous estimates.
Combining these developments into a calculation of the $p_T$-dependent nuclear modification factor and $v_2$ of inclusive $J/\psi$ production in
semi-central Pb-Pb collisions at the LHC, we find a good description of the experimental results by the ALICE collaboration. Our results thus resolve
the above-mentioned $v_2$ puzzle and imply the relevance of recombination processes for $p_T$'s of up to $\sim$8\,GeV.
\end{abstract}


\maketitle

{\it Introduction.---}
Heavy quarkonia have long been proposed as a probe of deconfinement in the hot QCD medium formed in ultrarelativistic heavy-ion
collisions (URHICs)~\cite{Rapp:2008tf,BraunMunzinger:2009ih,Kluberg:2009wc,Mocsy:2013syh,Rothkopf:2019ipj}. Based on their
spectroscopy in vacuum using the well-established Cornell potential, the main binding of charmonia is indeed generated by the linear
potential term commonly associated with the confining force. Lattice-QCD (lQCD) computations of the free energy of a static quark-antiquark
pair suggest that the confining force survives to temperatures significantly higher than the pseudo-critical temperature, $T_{\rm pc}$, of the
chiral phase transition. The initial suppression of charmonia in the hot medium of URHICs will then be followed by a subsequent ``regeneration"
once bound states are supported again, and inelastic reaction rates drive the charmonia yields toward their equilibrium values.
At which temperatures the different bound states (re-)emerge, and whether and how they reach equilibrium, are intensely debated questions at
present. The answers will depend on the microscopic interplay of the in-medium potential screening and the inelastic reaction rates.

\begin{figure*}[!tbh]
\includegraphics[width=0.495\textwidth]{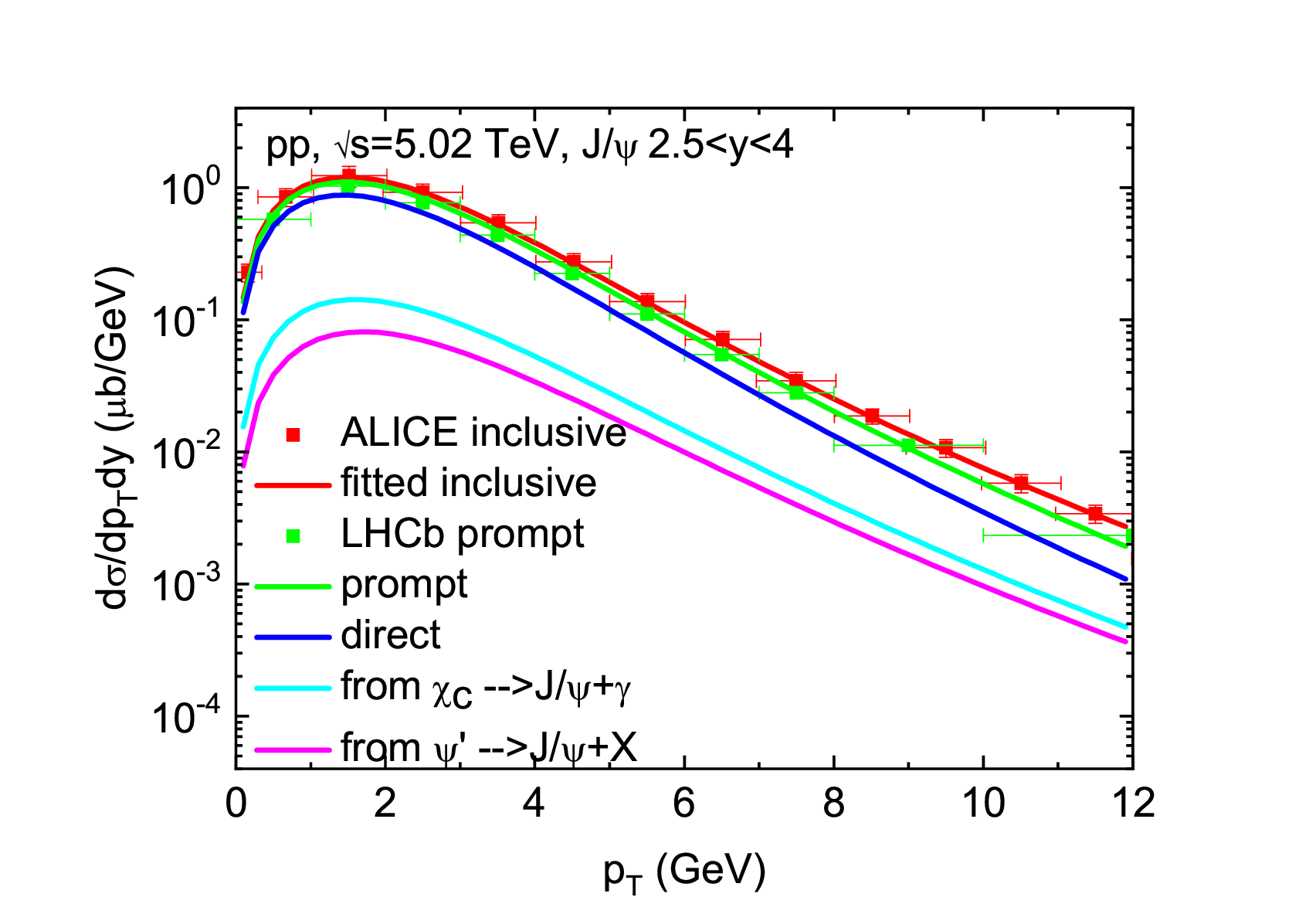}
\includegraphics[width=0.495\textwidth]{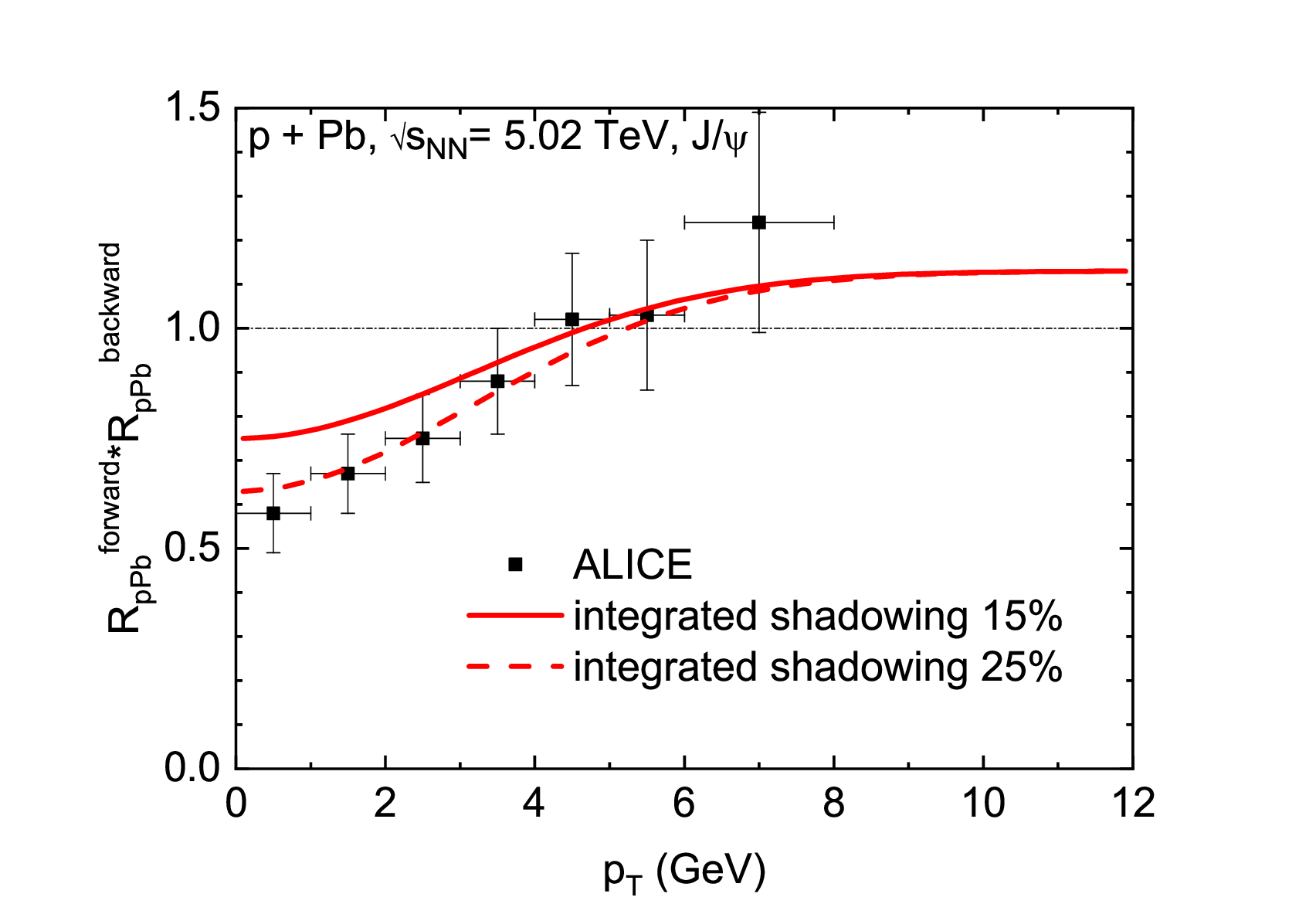}

\vspace{-0.4cm}

\caption{Left: Inclusive (red line) and prompt (green line: total, blue line: direct, cyan and pink line: $\chi_c$ and $\psi'$ feeddown)
$J/\psi$ spectra in 5.02\,TeV $pp$ collisions, compared to ALICE~\cite{ALICE:2021qlw} and LHCb~\cite{LHCb:2021pyk} data.
Right: Shadowing function (upper(lower) line for 15(25)\% total shadowing) compared to ALICE data~\cite{ALICE:2015sru} for the product of
forward and backward $R_{pA}$ in $p$-Pb collisions. Note that our transport approach~\cite{Du:2018wsj} generates an extra 10-15\%
suppression (not shown here) from hot-matter effects in $p$Pb collisions.}
\label{pp_dsigmadpTdy}
\end{figure*}
Heavy-ion collisions at the Super Proton Synchroton (SPS), Relativistic Heavy-Ion Collider (RHIC) and the Large Hadron Collider (LHC)
have provided a wealth of data on both charmonia and bottomonia. The former are rather strongly suppressed at SPS~\cite{NA50:2004sgj} and
RHIC energies~\cite{PHENIX:2006gsi,STAR:2016utm},
relative to the baseline from proton-proton ($pp$) collisions, but have shown a remarkable increase in their nuclear modification factor in
semi-/central Pb-Pb collisions at the LHC~\cite{ALICE:2016flj,ALICE:2019nrq,ALICE:2019lga,Bai:2020svs}. This rise has been predicted by both the
statistical hadronization model~\cite{Andronic:2006ky} and transport approaches~\cite{Yan:2006ve,Zhao:2011cv,Song:2011xi}, albeit the
predictions differ appreciably in detail. The presence of regeneration processes has been corroborated by experiment via the concentration of
the increased yield at low transverse momenta, $p_T\lsim 4$\,GeV, and an appreciable elliptic flow ($v_2$)~\cite{ALICE:2020pvw}, attributed
to the $v_2$ of recombining charm ($c$) quarks as found in $D$-meson observables~\cite{ALICE:2021rxa}.
However, predictions of transport model
calculations~\cite{Zhou:2014kka,Du:2015wha}, which give a fair description of the centrality and $p_T$-dependence of $J/\psi$ production in Pb-Pb
collisions at the LHC,  fall significantly short of the experimentally measured $v_2$ of $J/\psi$'s with transverse momenta $p_T\gsim 4 $\,GeV.
These calculations approximated the $p_T$ dependence of the regenerated $J/\psi$'s by a thermal blastwave ansatz (which implies thermalized
$c$- and $\bar{c}$-quark spectra), and also the elliptic flow of the primordial component was schematically (and conservatively) estimated to level
off at values of $\sim$2-3\% which underpredicts the data at high $p_T$. In the present paper we remedy these approximations by evaluating
the recombination processes with state-of-the-art $c$-quark distributions from relativistic Langevin simulations that give a fair description of open
heavy-flavor (HF) observables at the LHC~\cite{ALICE:2021rxa}. In this way we establish, for the first time, a quantitative connection between the
transport  of open and hidden charm particles in URHICs. We also revisit the $v_2$ calculations of the suppressed primordial $J/\psi$'s within the same
hydrodynamic model as used for the $c$-quark transport.

{\it Primordial $J/\psi$'s and their Suppression.--}
A common observable to characterize the hard production of particles in heavy-ion (AA) collisions, relative to that in proton-proton ($pp$)
collisions at the same energy, is the nuclear modification factor defined as
\begin{align}
R_{\rm AA}^{incl}(p_T)=\frac{dN^{\rm AA}_{incl}/dp_Tdy}{\langle T_{\rm AA}\rangle d\sigma^{pp}_{incl}/dp_Tdy}  \ ,
\label{raa}
\end{align}
where the numerator denotes the per-event particle spectrum for a given centrality selection, and $\langle T_{\rm AA}\rangle$ is the pertinent nuclear
thickness function~\cite{ALICE:2018tvk}. To construct the inclusive $J/\psi$ cross section in $pp$ collisions, $d\sigma^{pp}_{incl}/dp_Tdy$, we fit
pertinent 5\,TeV ALICE data~\cite{ALICE:2021qlw}, cf.~Fig.~\ref{pp_dsigmadpTdy}. Upon subtracting the $p_T$-dependent fraction
of non-prompt $J/\psi$ from $b$-hadron feeddown measured by LHCb at the same energy and forward rapidity~\cite{LHCb:2021pyk}, the resulting prompt $J/\psi$ spectra
agree with LHCb data~\cite{LHCb:2021pyk}, and are further decomposed into direct and feeddown contributions from $\chi_c$
and $\psi'$ decays using 7\,TeV LHCb data~\cite{LHCb:2012af,LHCb:2012geo}.

We implement nuclear shadowing using ALICE $J/\psi$ data~\cite{ALICE:2015sru} in $p$Pb collisions at rapidities 2.5$<$$|y|$$<$4 by
fitting the product of forward and backward $R_{p{\rm Pb}}$, cf.~Fig.~\ref{pp_dsigmadpTdy}. Our transport approach attributes
$\sim$10-15\% of the inclusive $J/\psi$ suppression in $p$Pb to hot-matter effects (mostly from feeddown)~\cite{Du:2018wsj}; thus, we
employ 15\% for nuclear shadowing as our baseline and take 25\% as an upper estimate for 20-40\% PbPb collisions.

\begin{figure*}[!t]
\includegraphics[width=0.495\textwidth]{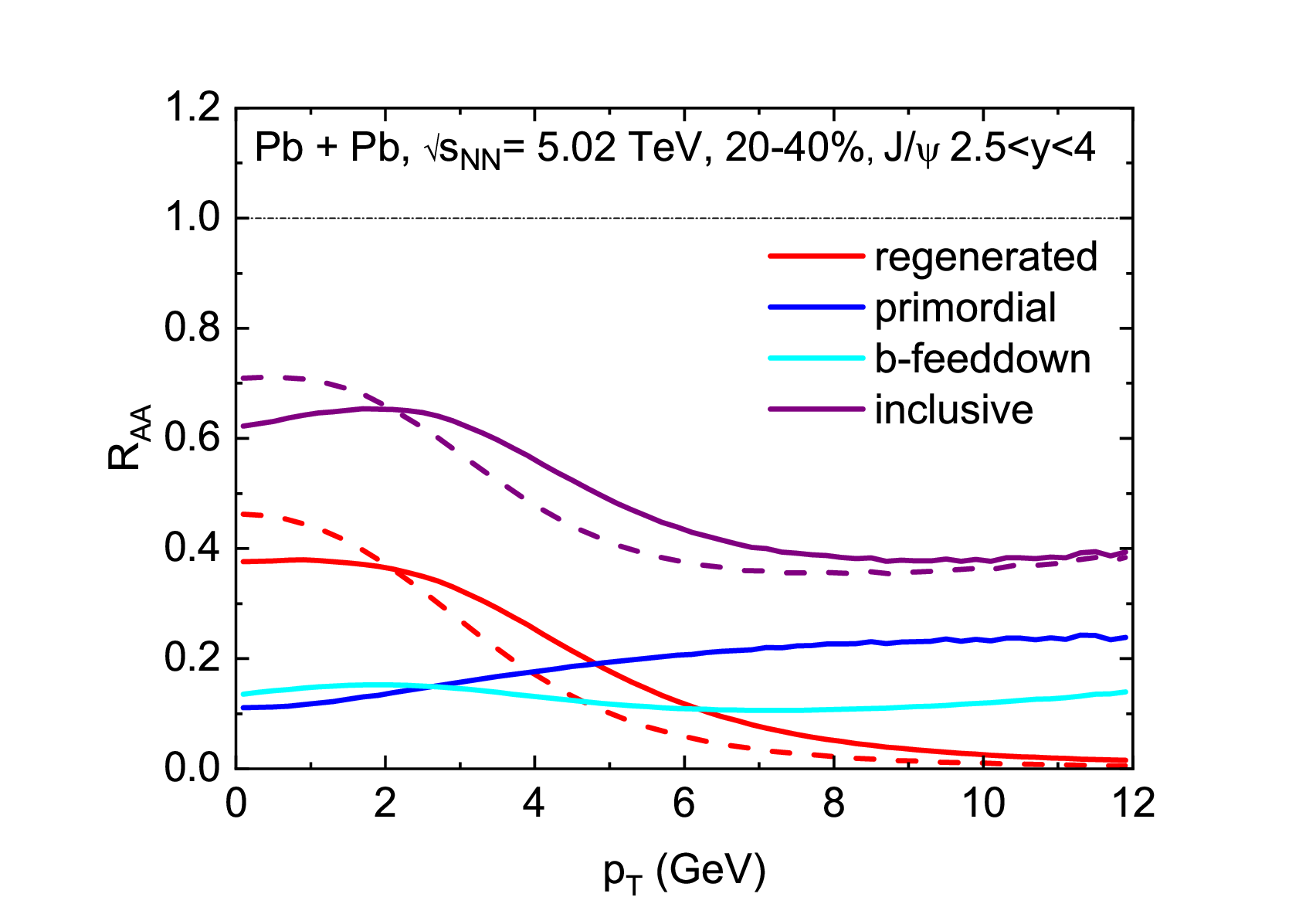}
\includegraphics[width=0.495\textwidth]{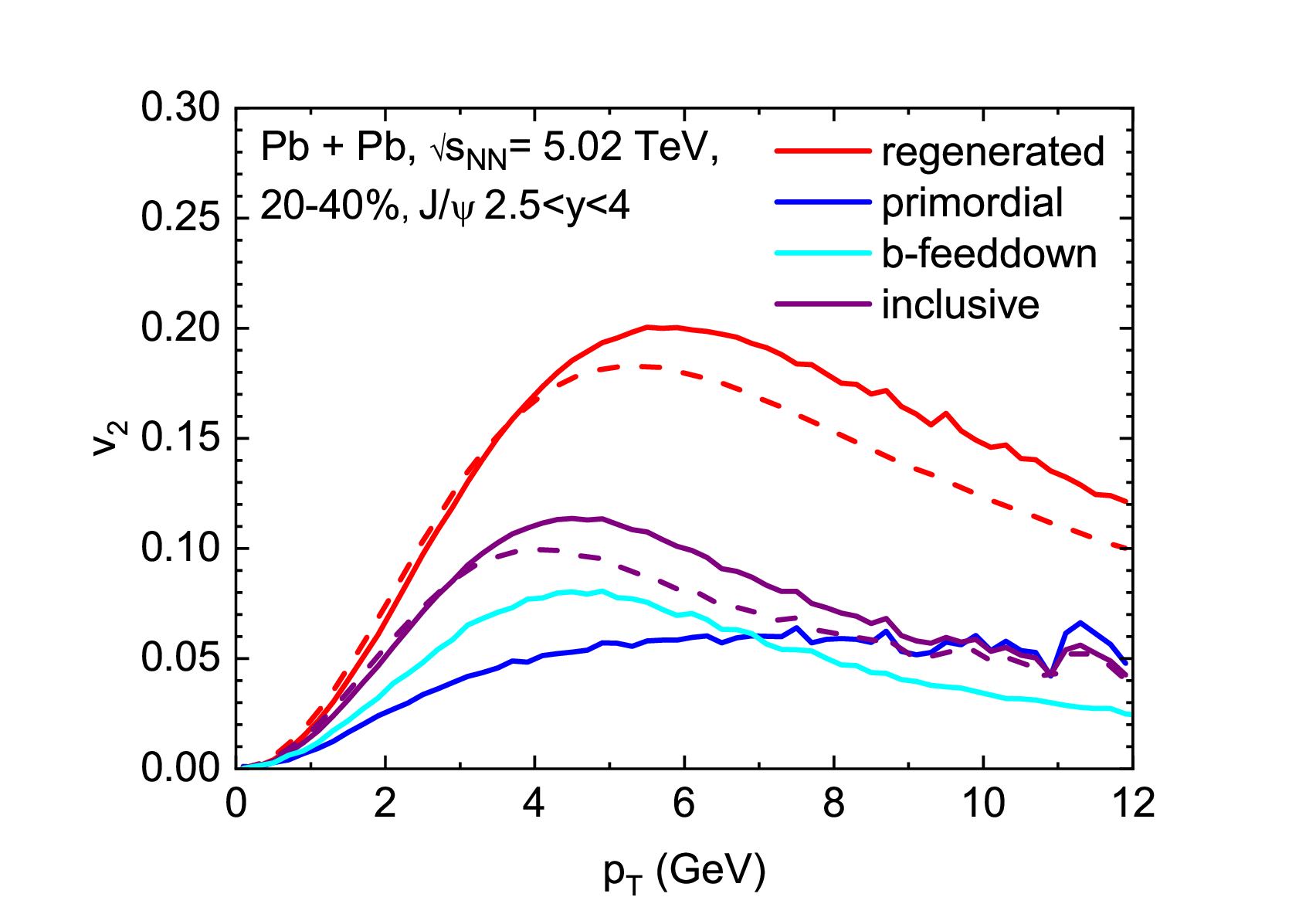}
\vspace{-0.7cm}
\caption{$J/\psi$ $R_{\rm AA}$ (left) and $v_2$ (right) in 20-40\% 5.02\,TeV PbPb collisions with 15\% integrated shadowing for inclusive production (purple lines; solid: with SMCs, dashed: without SMCs), regenerated (red lines; solid: with SMCs, dashed: without SMCs) and primordial components (blue lines), and bottom feeddown (cyan line).}
\label{fig_RAA_v2_different-components}
\end{figure*}

The suppression of primordial charmonia, $\Psi$=$J/\psi$, $\chi_c$ and $\psi'$, in Pb-Pb collisions utilizes the ``shadowed" $p_T$-differential
cross section and the spatial binary collision density function, $n_{\rm BC}(x,y)$, as initial conditions. Their straight-line motion through a smooth
hydrodynamic medium (which is tuned to light-hadron data at each centrality)~\cite{He:2019vgs} is simulated stochastically (\ie, event by event)
with quasifree dissociation rates and formation time effects from the transport approach of Ref.~\cite{Du:2015wha},
until the $\Psi$ enters a cell at the kinetic freezeout temperature (\eg, $T_{\rm kin}\simeq 110$\,MeV for 20-40\% central Pb-Pb collisions).
Following previous approaches, elastic rescattering is neglected, as it is parameterically suppressed.
At low $p_T$ the surviving prompt $J/\psi$'s shown in Fig.~\ref{fig_RAA_v2_different-components} are mostly direct $J/\psi$'s (the
$\chi_c$ and $\psi'$ are more strongly suppressed), while at high $p_T$ up to 40\% are from $\chi_c$ and $\psi'$ feeddown.
The magnitude and shape of the $p_T$ spectra are in fair agreement with the fireball calculations of Ref.~\cite{Du:2015wha}, but the
elliptic flow, levelling off at $\sim$5\% toward large $p_T$ in semicentral collisions (right panel of Fig.~\ref{fig_RAA_v2_different-components}),
is larger than earlier estimates of up to 2-3\%~\cite{Wang:2002ck,Zhao:2008vu,Zhou:2014kka,Du:2017qkv} using more schematic fireball
geometries.
In addition to the usual path-length dependence for the suppression along the short vs.~long axis of the almond-shaped nuclear overlap zone,
the hydrodynamic simulation features a larger density gradient along the short axis. The faster drop in density toward the boundary along the short axis
causes the outward propagating $\Psi$ to encounter smaller average densities than along
the long axis, implying on average smaller dissociation rates.

{\it $J/\psi$ Regeneration}.--
We compute the $p_T$-dependence of the regeneration component using the resonance recombination model (RRM)~\cite{Ravagli:2007xx}, which is
a 4-momentum conserving hadronization model based on the Boltzmann equation that recovers the correct equilibrium distribution of the produced hadron if
the underlying quark distributions are in equilibrium (also for moving media~\cite{He:2011qa}, including the $v_2$).
In its current implementation on a hydrodynamic hypersurface, the RRM does not predict the absolute norm of the produced hadrons. Therefore, we take recourse to our previous transport calculations for 3-momentum integrated $\Psi$ yields (with a $pp$ charm cross section of  $d\sigma_{c\bar c}/dy$=0.76\,mb based on recent ALICE data~\cite{ALICE:2021dhb} extrapolated to forward-$y$ using pQCD FONLL calculations; an analysis of the
uncertainty in this input for our results will be reported elsewhere). These are obtained from solving the kinetic-rate equation,
\begin{align}
\frac{dN_{\Psi}(\tau(T))}{d\tau}=-\Gamma_{\Psi}(T)[N_{\Psi}(\tau(T))-N_{\Psi}^{\rm eq}(T)] \ ,
\label{rate-eq}
\end{align}
in an expanding fireball medium, where the equilibrium limit, $N_{\Psi}^{\rm eq}(T)$, follows from relative chemical equilibrium of the time-evolving
charm content of the system ($c$-quarks in the quark-gluon plasma (QGP) and charm hadrons in hadronic matter, with total $c\bar{c}$ number conserved).
With the same inelastic reaction rates, $\Gamma_{\Psi}$, as used for the suppression calculation discussed above~\cite{Du:2015wha}, we can
utilize detailed balance to normalize the RRM yield to the regeneration yield of the rate equation.

\begin{figure}[!tbh]
\begin{minipage}[c]{0.98\linewidth}
\includegraphics[width=1.0\textwidth]{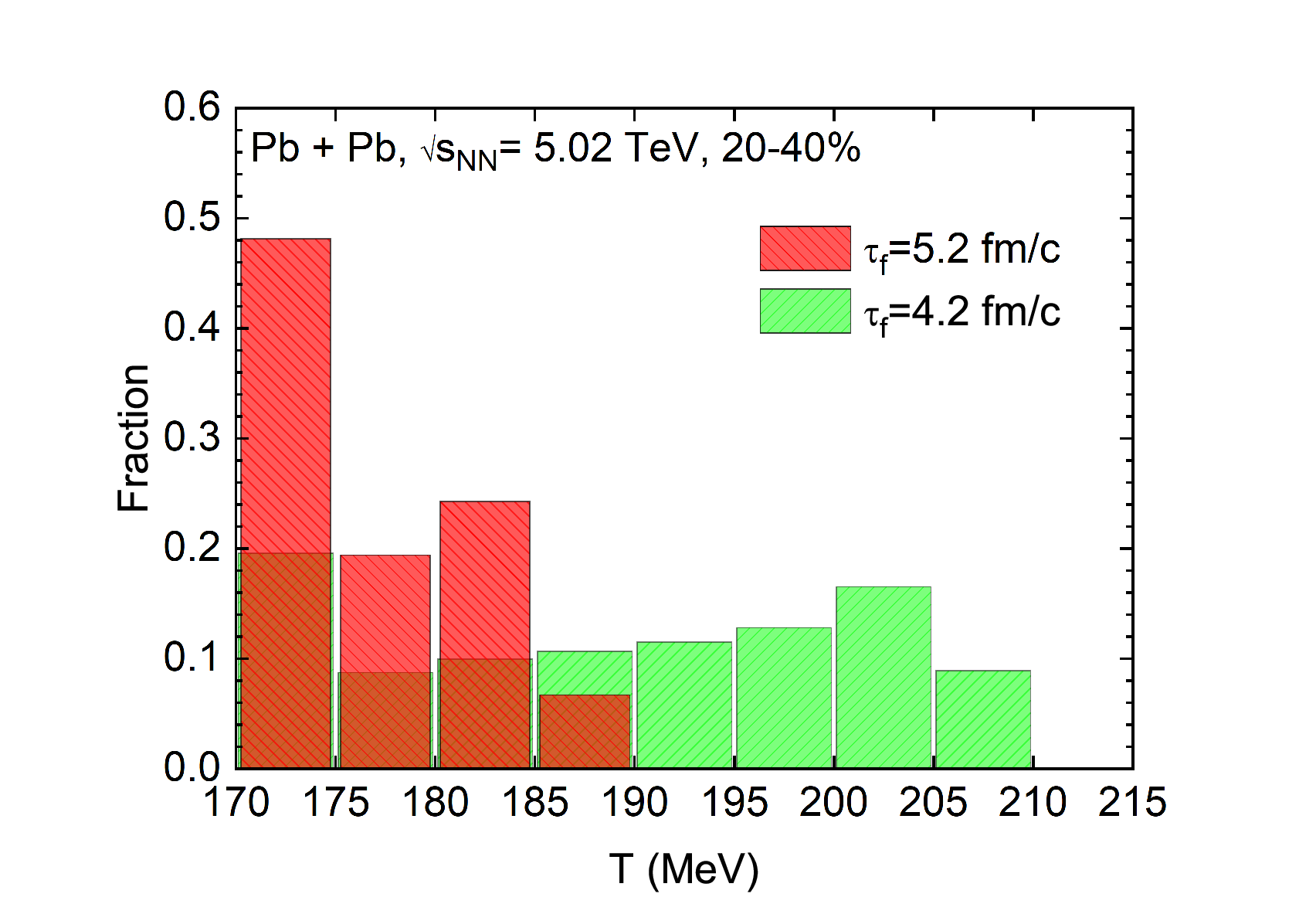}
\end{minipage}
\vspace{-0.2cm}
\caption{Temperature distribution of regenerated $J/\psi$'s computed with RRM on a hydrodynamic hypersurface at fixed proper time,
 $\tau_f$=5.2(4.2)~fm/$c$ (red (green) histogram).}
\label{fig_Tfo_distribution}
\end{figure}

The RRM $p_T$-spectrum of regenerated $\Psi$ states reads
\begin{align}
f_{\Psi}(\vec x,\vec p)=C_\Psi \frac{E_{\Psi}(\vec p)}{m_{\Psi}\Gamma_{\Psi}}
\int\frac{d^3\vec p_1 d^3\vec p_2}{(2\pi)^3}f_c(\vec x,\vec p_1)f_{\bar c}(\vec x,\vec p_2)
\nonumber\\
\times \sigma_{\Psi}(s)v_{\rm rel}(\vec p_1,\vec p_2)\delta^3(\vec p -\vec p_1-\vec p_2) \ ,
\label{rrm}
\end{align}
where $f_{\bar c,c}$ are the transported phase space distributions of $c$ and $\bar{c}$ quarks, and $v_{\rm rel}$ is their relative velocity.
In the current $2\to1$ formulation, hadronization is realized via a resonance cross section, $\sigma_{\Psi}(s)$, for $c+\bar c \to \Psi$,
taken of Breit-Wigner form with vacuum $\Psi$ pole masses and widths $\Gamma_{\Psi}\simeq 100$\,MeV (reflecting the dissociation
rates; variations by a factor of $\sim$2 have practically no effect on $p_T$-spectral shapes),
while the $C_\Psi$ ensures the normalization following from the rate equation.
The phase space distributions are taken from Langevin simulations of $c$-quark
diffusion in the QGP that result in a fair phenomenology of open HF observables at the LHC~\cite{He:2019vgs,ALICE:2021rxa}. In particular,
the RRM incorporates space-momentum correlations (SMCs)~\cite{He:2019vgs}, here between $c$ and $\bar{c}$ quarks from event-by-event
Langevin simulations.
In the rate equation approach, the production of charmonia occurs over a finite temperature range~\cite{Zhao:2011cv}.
In analogy to the procedure adopted before where the $J/\psi$ blastwave spectra were evaluated at an {\em average} temperature of
$T$=180\,MeV~\cite{Du:2015wha}, we here evaluate the RRM on a hydrodynamic hypersurface at a constant longitudinal proper time,
$\tau_f\simeq5.2$\,fm/$c$, for semi-central 5.02\,TeV Pb-Pb collisions, yielding a range of average temperatures of $T$=170-190\,MeV,
cf.~the red histogram in Fig.~\ref{fig_Tfo_distribution}.
Since the temperature window for regeneration is related to the dissociation temperature of the various $\Psi$ states, we also consider a 20\% earlier
freezeout time, $\tau_f\simeq4.2$\,fm/$c$, resulting in a significantly higher range of average regeneration temperatures, cf.~the green histogram in Fig.~\ref{fig_Tfo_distribution}.

\begin{figure*}[!thb]
\begin{minipage}[c]{0.5\linewidth}
\vspace{-0.5cm}
\hspace{-0.4cm}
\includegraphics[width=1.0\textwidth]{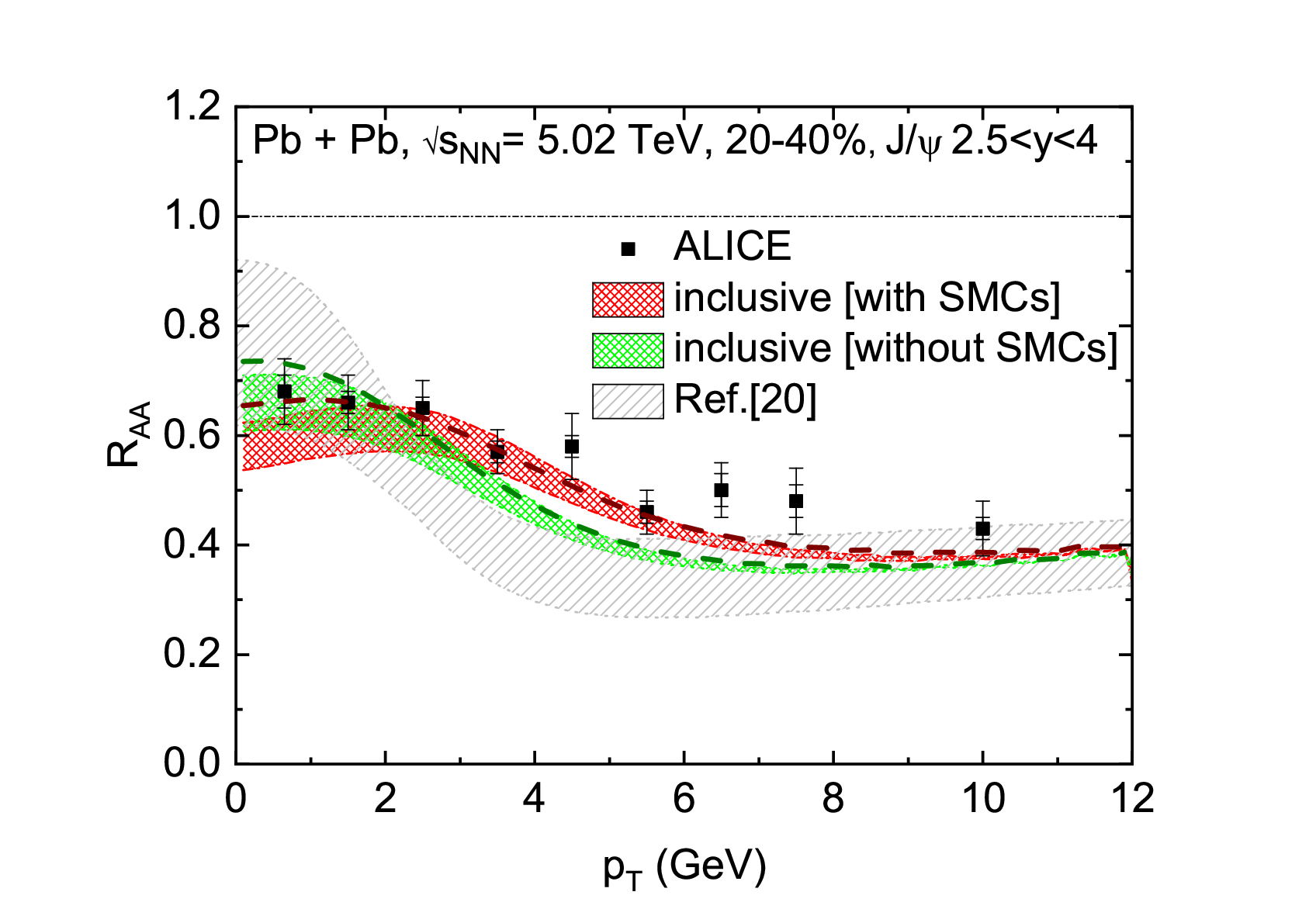}
\end{minipage}
\begin{minipage}[c]{0.49\linewidth}
\vspace{-0.5cm}
\hspace{-0.4cm}
\includegraphics[width=1.0\textwidth]{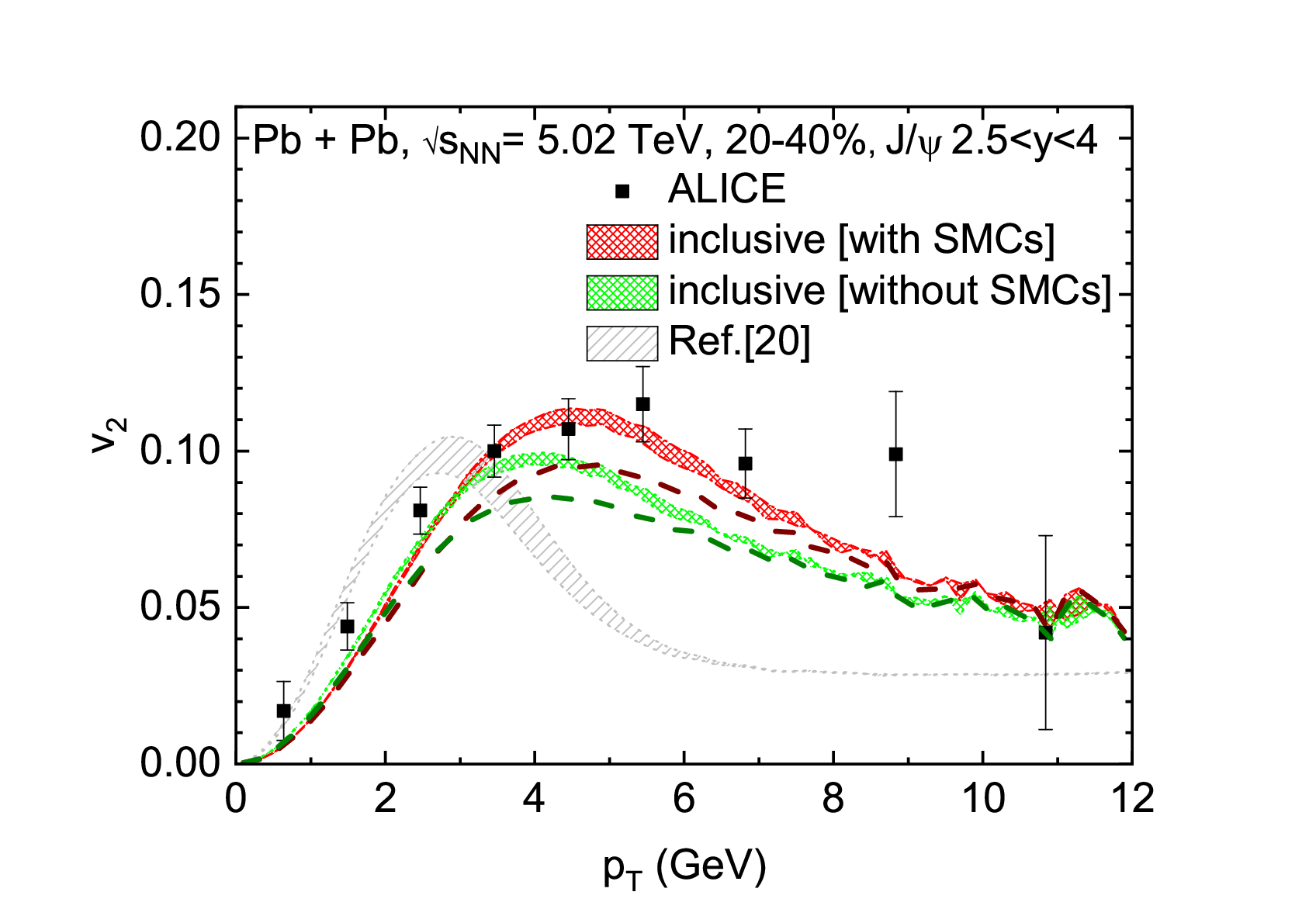}
\end{minipage}

\vspace{-0.3cm}

\caption{Inclusive-$J/\psi$ $R_{AA}$ (left) and $v_2$ (right) in 5.02\,TeV Pb-Pb collisions at the LHC, compared to ALICE
data~\cite{ALICE:2019lga,ALICE:2020pvw}. The red (green) bands employ $c$-quark spectra in RRM
evaluated at $\tau_f$=5.2~fm/$c$ with(out) SMCs, where the widths reflect a 15-25\% charm and charmonium shadowing range.
The dashed lines are for $\tau_f$=4.2~fm/$c$ with 15\% shadowing (brown (dark-green) dashed: with(out) SMCs). Inclusive results from
previous calculations~\cite{Du:2015wha,ALICE:2019lga} (grey bands) are shown for comparison.}
\label{fig_RAA-v2}
\end{figure*}
Compared to earlier thermal-blastwave approximations~\cite{Zhou:2014kka,Du:2015wha,Andronic:2015wma}, the RRM $J/\psi$
spectra in 20-40\% 5.02\,TeV Pb-Pb collisions with transported $c$-quark distributions are appreciably larger for $p_T \gsim3$\,GeV.
The inclusion of SMCs is also quite noticeable, enhancing the result without SMCs for $p_T$$\gsim$3\,GeV, \eg,
by a factor of $\sim$2 at $p_T$$\simeq$6\,GeV, where the recombination yield is still significant, see Fig.~\ref{fig_RAA_v2_different-components} left,
especially since its $v_2$ is much larger than the primordial one, reaching up to $\sim$20\%,  cf.~Fig.~\ref{fig_RAA_v2_different-components}
right. Also here the SMCs lead to an increase, by ca.~10-20\% for $p_T$$>$4\,GeV. For the inclusive $v_2$, the
extended reach of regeneration due to the SMCs causes a relative increase of up to 30\% for $p_T$$\simeq$5-7\,GeV.

{\it Non-Prompt $J/\psi$}.--
The final component of inclusive $J/\psi$ yields as measured by ALICE is the ``non-prompt"
contribution from weak decays of $b$-hadrons. We also utilize transported $b$-quark spectra with subsequent
resonance recombination into $B$-mesons as computed earlier~\cite{He:2014cla} (we do not resolve
$\Lambda_b$ baryons, although the total $b$-decay contribution is encoded in the fit to $pp$ data). These are decayed into $J/\psi$'s using the
momentum distributions measured in Ref.~\cite{BaBar:2002zxr}. The absolute differential yield of non-prompt $J/\psi$ is determined
in analogy to prompt feeddown contributions, based on the $pp$ fit constructed in Fig.~\ref{pp_dsigmadpTdy} left.

{\it Inclusive $J/\psi$}.--
The final results for the inclusive $J/\psi$ $R_{\rm AA}(p_T)$ in 5.02\,TeV Pb-Pb collisions follow by adding the direct-suppressed and RRM components
including their feeddown contributions as well as the non-prompt part from transported $b$ quarks, in the numerator and dividing by the denominator
as described following Eq.~({\ref{raa}). The $v_2(p_T)$ is obtained by a weighted sum of the individual components.  The theoretical error bands
represent the 15-25\% shadowing effect on the $p_T$-dependent input distributions of primordial $J/\psi$'s
and $c$-quarks. We find a rather good description of the ALICE data for both $R_{\rm AA}$ and $v_2$, cf.~Fig.~\ref{fig_RAA-v2}, thus resolving the
disagreement with previous transport model results. At intermediate $p_T$, the harder off-equilibrium $c$-quark spectra (as well as SMC effects)
much extend the reach of the regeneration component, significantly exceeding previously employed blastwave spectra. While the net increase in the
$R_{\rm AA}$ around $p_T\simeq4$\,GeV is moderate, its regeneration fraction is much enhanced which, in turn, is the key effect in increasing
the inclusive $v_2$.
In addition, the considerable increase in the primordial $v_2$ at high $p_T$, from $\sim$3 to $\sim$5\%, is quite relevant for $p_T\gsim 5$\,GeV.
We illustrate the uncertainty of the regeneration temperature window by a 20\% reduction of the longitudinal proper time for the transported $c$-quark
spectra ($\tau_f=4.2$\,fm/$c$), cf.~the dashed lines in Fig.~\ref{fig_RAA-v2} (with a 15\% integrated shadowing, to
be compared to the upper edges of the bands for $\tau_f=5.2$\,fm/$c$). The $R_{\rm AA}$ is little affected by this change (mostly via a
small enhancement at low $p_T$), but the $v_2$ is reduced by up to 15-20\%. This is appreciable and may be used in future analysis to more
precisely constrain the regeneration window.
Since in our present treatment the regeneration temperatures represent {\em average} values, an explicit account of the time dependence of the
$c$-quark phase space distributions, in connection with the underlying $T$-dependent dissociation rates and melting temperatures of
different $\Psi$ states, will be in order.

{\it Summary.--}
We have calculated the nuclear modification factor and elliptic flow of inclusive $J/\psi$ production in heavy-ion collisions at the LHC. The
main theoretical development was the implementation of state-of-the-art $c$-quark phase space distributions into the regeneration process of
charmonia as obtained from a strongly coupled transport approach for HF diffusion that describes open-charm observables at the LHC. Both
momentum and spatial dependences of these distributions (through off-equilibrium spectra and space-momentum correlations, respectively) were
found to extend the relevance of recombination processes up to transverse momenta of about 8\,GeV in semi-central 5.02\,TeV Pb-Pb collisions.
In addition, an explicit calculation of the suppression of primordially produced charmonia leads to a significantly higher $v_2$ (caused by the path-length
dependent dissociation) than previously estimated. We also utilized transported $b$-quark distributions to calculate the non-prompt portion of the
inclusive $J/\psi$ yields. Putting all these together, a good description of the experimental data for the $R_{\rm AA}$  and $v_2$ of inclusive
$J/\psi$'s emerged, thereby resolving discrepancies found in earlier transport model calculations.
Our calculations exhibit significant sensitivity to the (average) production time of the $J/\psi$ through its collectivity imprinted by the regeneration
processes, where early production is disfavored by the magnitude of the $v_2$ and late production (with a larger radial flow) by the development
of a characteristic ``flow bump" in the $R_{\rm AA}$ at low $p_T$ which is limited by current data. Future precision analysis, aided by statistical
tools, will likely provide further constraints on the dissociation temperatures of the various charmonia.
Our study straightforwardly extends into the bottom sector (where previous calculations for $Y$ production exist within a weak-coupling
approximation~\cite{Blaizot:2017ypk,Yao:2020xzw}), to $B_c$ mesons, and exotic quarkonia. While for $Y$ mesons regeneration contributions
are expected to be moderate~\cite{Du:2017qkv}, they are presumably more important for the $B_c$ and exotica~\cite{Wu:2020zbx},
especially relative to their production in $pp$. Also for these observables our framework can improve the understanding of the
pertinent dissociation kinetics. Work in this direction is in progress.

{\it Acknowledgments.--}
This work was supported by NSFC grant 12075122 and  U.S.~NSF grant no.~PHY-1913286. MH acknowledges helpful
discussions with X. Du.

\end{document}